\documentclass[pra,aps,superscriptaddress,twocolumn,showpacs,nofootinbib]{revtex4-1}

\usepackage{placeins}
\usepackage{amsfonts, amsmath,amssymb,amsthm,bm}
\usepackage{mathrsfs}
\usepackage{graphicx}
\usepackage[usenames,dvipsnames]{color}
\usepackage{palatino}
\usepackage{fancyhdr}
\usepackage{mathrsfs}
\usepackage{multirow}
\usepackage{hyperref}
\usepackage{bigints}
\usepackage{framed}
\definecolor{shadecolor}{gray}{0.9}

\hypersetup{
   colorlinks=true,
   linkcolor=blue,
   citecolor=blue,
   urlcolor=Violet
}

\theoremstyle{plain}

\theoremstyle{definition}

\newcommand{\IQOQI}{Institute for Quantum Optics and Quantum Information,\\ Austrian Academy of Sciences, Boltzmanngasse 3, A-1090 Vienna, Austria}
\newcommand{\Peri}{Perimeter Institute for Theoretical Physics, 31 Caroline Street North, Waterloo, ON N2L 2Y5, Canada}

\begin{document}
\title{Undecidability and unpredictability: not limitations, but triumphs of science}
\author{Markus P.\ M\"uller}
\affiliation{\IQOQI{}}
\affiliation{\Peri{}}

\begin{abstract}
It is a widespread belief that results like G\"odel's incompleteness theorems or the intrinsic randomness of quantum mechanics represent fundamental limitations to humanity's strive for scientific knowledge. As the argument goes, there are truths that we can never uncover with our scientific methods, hence we should be humble and acknowledge a reality beyond our scientific grasp. Here, I argue that this view is wrong. It originates in a naive form of metaphysics that sees the physical and Platonic worlds as a collection of things with definite properties such that all answers to all possible questions exist ontologically somehow, but are epistemically inaccessible. This view is not only a priori philosophically questionable, but also at odds with modern physics. Hence, I argue to replace this perspective by a worldview in which a structural notion of \emph{`real patterns', not `things'} are regarded as fundamental. Instead of a limitation of what we can know, undecidability and unpredictability then become mere statements of \emph{undifferentiation of structure}. This gives us a notion of realism that is better informed by modern physics, and an optimistic outlook on what we can achieve: we can know what there is to know, despite the apparent barriers of undecidability results.
\end{abstract}

\date{November 13, 2020}

\maketitle

\section{The pessimistic view}
\label{SecPessimistic}
The early 20th century has given us insights into mathematics, physics, and computer science that seemed to shatter our hope for unlimited progress of scientific knowledge. In 1931, G\"odel published his famous incompleteness theorems~\cite{Goedel}, implying that every sufficiently complex consistent axiomatic system contains statements that are true but unprovable within the system. An information-theoretic version of this insight is Turing's proof of the unsolvability of the halting problem~\cite{Turing}: there is no algorithm that could, in all instances and in finite time, decide whether another specified computation will eventually halt or run indefinitely. At about the same time, the discovery of quantum physics has led us to the insight that Nature seems to be intrinsically random: even with maximal knowledge of the current state of the world, it is impossible to predict future events with certainty~\cite{EPR}.

At first glance, these insights seem to point at limitations of science, suggesting an attitude of humility and disappointment. Before these results, there was David Hilbert's program to reduce all of mathematics to a finite, complete, provably consistent set of axioms~\cite{HilbertSEP}. And there was Pierre-Simon Laplace's famous declaration~\cite{Laplace} of the possibility of a ``demon'', able to predict all of the future with certainty given sufficient physical data. The new insights were in stark contrast to these declarations, showing that the hopes of Hilbert, Laplace and others were misplaced. Does this mean that mathematics and physics are, as scientific disciplines, intrinsically deficient in some sense?

At least this is the way that these theorems are often portrayed, both in popular and in academic accounts. Regarding G\"odel's theorems, Wikipedia~\cite{WikiGoedel} claims: \emph{``These results set a limit in principle to mathematics: not every mathematical theorem can be formally derived or disproved from the axioms of some area [...] of mathematics.''} This proposed limitation of mathematics is often contrasted to an alleged omnipotence of the human mind, leading to a class of ``anti-mechanist'' arguments: \emph{``There have been repeated attempts to apply G\"odel's theorems to demonstrate that the powers of the human mind outrun any mechanism or formal system.''}~\cite{GoedelSEP}. Philosopher John R.\ Lucas~\cite{Lucas} claims that \emph{``given any machine which is consistent and capable of doing simple arithmetic, there is a formula it is incapable of producing as being true [...] but which we can see to be true.''} (cited in~\cite{GoedelSEP}).

If, on the other hand, one gives up on the idea that the human mind is in some specific sense more powerful than any mechanism, then it may be tempting to read G\"odel's theorem as a fundamental epistemic restriction for humanity. This view is vividly expressed, for example, by Driessen and Suarez~\cite{DriessenSuarez}:

\emph{``In this book, recent mathematical theorems are discussed, which show that man never will reach complete mathematical knowledge. Also experimental evidence is presented that physical reality will always remain partially veiled to man, inaccessible to his control. It is intended to provide, in the various contributions, the pieces of a puzzle which restore the possibility of a natural, intellectual access to the existence of an omniscient and omnipotent being.''}

As the quotation indicates, there is a widespread view of quantum physics which regards its statistical character as a symptom of incompleteness. This view is defended, for example, by proponents of de Broglie-Bohm theory~\cite{Bohm}, a nonlocal hidden-variable interpretation of quantum mechanics. According to this view, there exists a deterministic underlying reality, and particles have well-defined positions at any time. However, the predictions of quantum mechanics are probabilistic due to fundamental uncontrollable disturbances. It is this unavoidable lack of experimental control that is ultimately responsible for the apparent randomness of measurement outcomes. According to this view, quantum theory's statistical character is thus most naturally interpreted as manifesting some fundamental epistemic restriction.

So has science, in the problems described above, hit an impenetrable barrier? Let us gain some intuition by looking at a problem where humanity seemed to hit a barrier for about two thousand years, before the problem became finally dissolved.

\section{On axiomatic theories and\\ structural differentiation}
\label{SecMath}
Around 300 BC, in his treatise \emph{Elements}, Euclid formulated a set of axioms and postulates that were supposed to capture the essence of geometry~\cite{Euclid}. One of these principles seemed less self-evident than the others and was hence standing out: Euclid's fifth postulate, the \emph{parallel postulate}. Could this postulate perhaps be proven as a consequence of the others? This hope was the source of a twenty-centuries-long search for a proof, for a logical deduction of the uniqueness of any parallel line through any given external point from simpler assumptions.

In the 19th century, this search finally came to an end. The discovery of nonstandard geometry showed that the parallel postulate cannot logically follow from the others. When it became gradually clear that (what we call today) elliptical and hyperbolic geometries are consistent theories -- and these theories satisfy all other principles \emph{except} for the fifth postulate -- the parallel postulate changed its status from an apparent necessity to a choice to be made.

\begin{figure}[hbt]
\begin{center}
\includegraphics[width=.4\textwidth]{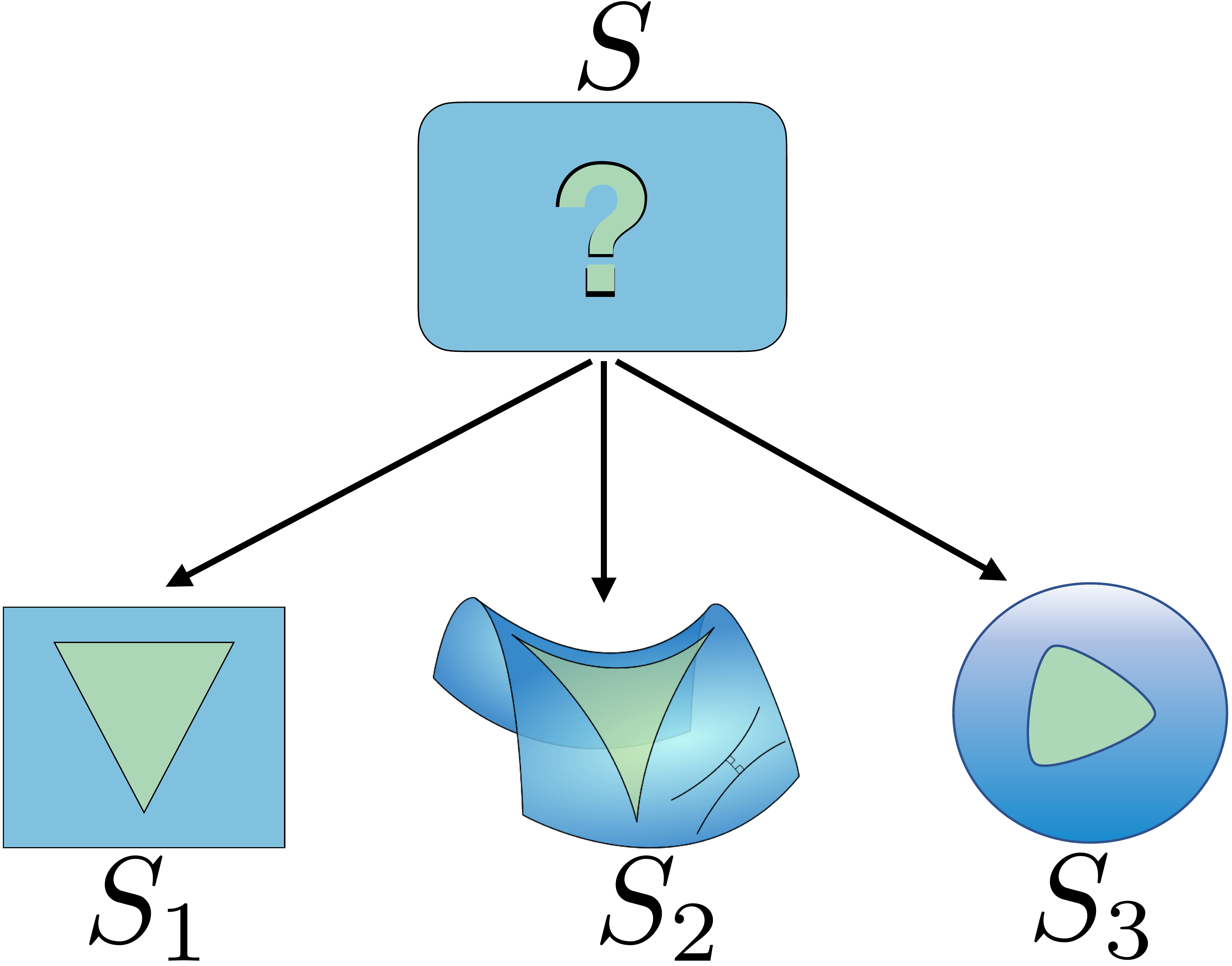}
\caption{Absolute geometry and its differentiations.}
\label{fig_geometry}
\end{center}
\end{figure}

Denote by $T$ the totality of all of Euclid's axioms and postulates (or rather, of Hilbert's more rigorous reformulation~\cite{Hilbert}) \emph{except} for the parallel postulate -- we can see it as a formal system, or \emph{theory}. But what is this theory $T$ about? It describes geometric objects -- points, lines, circles, and more -- and their relations. It allows us to prove many interesting statements about these objects (such as: an exterior angle of a triangle is larger than either of the remote angles), but it leaves some questions undecided that seem natural to ask (is the sum of the angles in a triangle equal to $180^\circ$?). This theory is sometimes called \emph{absolute geometry} (see~\cite{Greenberg} for details).

If we add the parallel postulate to $T$, we obtain another theory: $T_1$, familiar \emph{Euclidean geometry}. On the other hand, we can add suitable modifications of the parallel postulate to $T$, and obtain $T_2$ and $T_3$: \emph{hyperbolic} and \emph{elliptic geometry}.

When we talk about a theory in this way, we mean a systems of axioms equipped with formal rules to generate theorems, formulated in some language. However, we typically have a mental picture of \emph{the things that the theory talks about} -- the ``meaning'' of the language. Euclidean geometry $T_1$, for instance, is typically not envisioned as an abstract language game, but vividly depicted as talking about \emph{geometric structure}: geometric objects embedded in a plane. The standard mathematical description of this idea is to say that a theory can have a \emph{model}~\cite{Marker}: a set (say, the two-dimensional plane) equipped with distinguished elements (such as points, lines, circles), functions, and relations (such as incidence or congruence) such that the theorems of the theory are true when understood as talking about these elements.

For what follows, we can take a somewhat different perspective which is implicitly shared by many practicing mathematicians, but rarely explicitly communicated. Let us stipulate the following informal definition.

\begin{shaded}
\textbf{Definition.} A \emph{structure} $S$ is whatever is described by a consistent theory $T$.
\end{shaded}

For example, the structure $S_1$ described by theory $T_1$ corresponds to the objects and relations of Euclidean geometry -- points, lines,  their incidences and congruences, and several others. More formally, we can define a structure $S$ as the collection of all models of its theory $T$.

Figure~\ref{fig_geometry} gives a sketch of the geometric structures mentioned above. Euclidean, hyperbolic and elliptic geometries $S_1$, $S_2$ and $S_3$ are \emph{more differentiated structures} than absolute geometry $S$: they have all the properties of $S$, plus some additional properties. On the other hand, we can regard $S$ as the collection of all its differentiations $S_1\cup S_2 \cup S_3$, because every model for $T_1$, $T_2$ or $T_3$ is also a model for $T$.

\begin{shaded}
\textbf{Definition.} Structure $S'$ is \emph{more differentiated} than structure $S$ if its corresponding theory $T'$ is an extension of the theory $T$ --- that is, if $T'$ contains the same formal rules and axioms as $T$, plus additional axioms. This implies that all models for $T'$ are also models for $T$.

Consequently, a structure is equal to the collection of all its differentiations.
\end{shaded}

The theory $T$ describing absolute geometry is incomplete. That is, the question $Q$ \emph{``is there more than one line through any given point parallel to another line?''} is undecidable within $T$: neither the affirmation nor the negation of this question can be proven in $T$. However, we can still regard the corresponding structure $S$ as a perfectly valid ``thing'' in some sense. It is simply the case that some questions (like $Q$) that we may ask about this ``thing'' \emph{don't have answers}. An answer exists for differentiations $S_1$, $S_2$ and $S_3$ of $S$ (no, yes, and no), but no answer to this question exists for $S$. And this is how it \emph{must} be: since $S$ is the collection of all its differentiations, neither affirmation nor negation\footnote{Note that this does not violate the law of the excluded middle. The statement $A\vee\neg A$ (for $A$ the answer to $Q$) is still true for $S$, precisely because it holds for all its differentiations.} of $Q$ can be true for $S$. In this sense, undecidability of $Q$ in theory $T$ simply refers to the fact that the corresponding structure $S$ is \emph{undifferentiated} with respect to $Q$. It is a bit like stem cells of the human embryo, which are undifferentiated in the sense that the question ``what type of cells will these become?'' does not (yet) have an answer.

But should we really regard $S$ as a valid ``structure'' if it has these ``holes'' in its catalogue of properties? Isn't $S$ a defective thing, since the corresponding theory $T$ is defective, i.e.\ incomplete?

If we decided to throw out $S$ on the basis of $T$'s incompleteness, then we would quickly run out of interesting mathematical structures. This is precisely the content of \emph{G\"odel's first incompleteness theorem}:
\begin{shaded}
\textit{``Any consistent theory $T$ within which a certain amount of elementary arithmetic can be carried out is incomplete; i.e., there are statements of the language of $T$ which can neither be proved nor disproved in $T$.''}~\cite{GoedelSEP}
\end{shaded}
Thus (see also~\cite[Sec.\ 2.6]{GoedelSEP}), we can obtain new theories $T'$ and $T''$ by adding an unprovable statement \emph{or its negation} as a new axiom to $T$. This means that the corresponding structure $S$ has (at least) two inequivalent differentiations, $S'$ and $S''$ -- similarly as absolute geometry has elliptic and hyperbolic (and Euclidean) geometries as differentiations. Calling a structure ``interesting enough'' if its theory $T$ admits the necessary amount of arithmetic to be formalized for G\"odel's theorem to apply, we arrive at the following consequence:
\begin{shaded}
\textbf{Theorem.} Every interesting enough structure has several inequivalent differentiations.
\end{shaded}
Intuitively, and perhaps traditionally, we tend to think of the mathematical world as consisting of ``mathematical objects'' with a catalogue of statements that are ontologically either true or false. For example, we may believe that there is something called ``the natural numbers'', $\mathbb{N}$, a well-established ``thing'' (after all, formalized as a \emph{set}) that somehow ``sits there'', waiting for our mathematical tools to discover all of its properties and to prove all of its true theorems. Since mathematicians are only human, as this informal argument goes, all they can do is resort to theories $T$ that try to capture the essence of $\mathbb{N}$ (such as the Peano axioms), and to use these theories to prove results about $\mathbb{N}$. According to this view, G\"odel's first incompleteness theorem is bad news: it implies that there will always be statements about $\mathbb{N}$ that are true, but that cannot be proven by our best theory.

The terminology established above allows us to take a different perspective on the Platonic world. If we visualize the mathematical world as consisting of structure in the above sense, then G\"odel's first incompleteness theorem attains a quite different, more optimistic interpretation: proving the undecidability of a statement is not a certificate of principled human fallibility, but a deep insight into the \emph{existence of several distinct differentiations of some structure}. It is not a fundamental limit to what we can know, but a precious piece of knowledge about a \emph{non-property} of the structure that we have discovered.

\section{The physical world: every thing must go}
\label{SecPhys}
Modern physics, as I will now argue, informs us that a similar move should be considered regarding our \emph{physical} world. Consider the historical notion of the \emph{luminiferous aether}. For several centuries, it was believed that light waves need a material medium for propagation, similarly as water waves. It was therefore natural to ask: \emph{What are the properties of aether? How can we experimentally verify its existence and its properties?}

The historical course of events is well-known. After the exploration of light revealed more and more implausible properties of aether, the Michelson-Morley experiment and the subsequent development of Special Relativity has finally led the physics community to abandon this notion. This turn of events meant that the inability to answer the questions above (what are the properties of aether?) was not due to experimental limitations, but due to the fact that \emph{the questions have no answers}. In other words: the questions were not solved, but \emph{dissolved}.

The aether and its properties is by far not the only problem of physics with the final fate of dissolution. Consider the following question:\\

\emph{Did events $A$ and $B$ happen at the same time?}\\

This is a very natural question with many highly relevant instances. For example, \emph{did homo sapiens and homo erectus inhabit Southeast Asia at the same time?} Questions like this are immensely important for understanding our human ancestry~\cite{Swisher}. \emph{Did the suspect arrive at his hotel room at the same time that the victim was killed in the bar?} The answer may well determine whether the suspect is sent to prison.

Even though we can (hopefully) answer the question with enough effort in the cases just described, Special Relativity tells us that we cannot obtain an answer in all instances. But is this due to a limitation of our experimental abilities? No. According to Special Relativity, it is because \emph{the above question doesn't have an answer}. It is only that in the cases of interest (such as the two just described), the question is implicitly asked relative to a predetermined frame of reference. Based on Newtonian mechanics, we thought that a general, absolute answer to this question \emph{should} always exist, we we have found that it doesn't.

The dissolution of questions tends to provoke considerable resistance. This is even true for well-established insights like the relativity of simultaneity, as the following quotation by philosopher and logician Peter Geach (\cite{Geach}, cited in~\cite{Ladyman}) demonstrates: \emph{``[...] `at the same time' belongs not to a special science but to logic. Our practical grasp of this logic is not to be called into question on account of recondite physics [...] A physicist who casts doubt upon it is sawing off the branch he sits upon''}.

If such well-established instances of dissolution like the absence of absolute simultaneity provoke resistance, then it should not come as a surprise that such hesitation is particularly strong in the context of our second revolution of modern physics: quantum theory.

Quantum theory claims that there are questions that we may be interested in asking, but that we will never be able to answer, no matter what heavy artillery of physical methods we roll out. If we decide to prepare a quantum state in the superposition $|\psi\rangle=\frac 1 {\sqrt{2}}(|0\rangle+|1\rangle)$ and measure, we may be interested to know beforehand whether we will obtain outcome $|0\rangle$ or $|1\rangle$. But if quantum theory is correct, then this desired prediction is impossible. It is \emph{as rock solid impossible}~\cite{ColbeckRenner} as it is impossible to absolutely decide, in the regime of Special Relativity, whether two events $A$ and $B$ happened simultaneously.

Let us reformulate this observation of unpredictability. The question that turns out to be unanswerable is arguably best characterized as follows:\\

\emph{What is, at some given moment, the \emph{actual configuration} of the world?}\\

There are different ways to understand this question, depending on what we mean by an ``actual configuration''. In the foundations of quantum mechanics, this notion is often understood in a particular way: as a collection of values of ordinary variables that resemble what John Bell has called ``beables''~\cite{BellBeables}. If such beables exist, and if they determine the outcomes of quantum measurements, then it is in principle impossible for us to get to know them all. These hypothetical ``hidden variables'' are not only epistemically inaccessible, but they also have properties that seem implausible. For example, by Bell's theorem, these hidden variables must be \emph{nonlocal} in some sense; the way that they manifest themselves in measurements on entangled quantum states must necessarily involve superluminal signalling, but this signalling is miraculously washed out (``fine-tuned''~\cite{WoodSpekkens}) so that it does not show up in the workings of our devices.

De Broglie-Bohm theory consists precisely of an attempt to answer the above question with an interpretation of the ``actual configuration'' of the world as just described. But also some proponents of $\psi$-epistemic interpretations~\cite{Leifer,Spekkens} \emph{(what kind of hidden variables with what kind of causal structure give rise to quantum states as states of knowledge?)} or QBism~\cite{Fuchs} \emph{(what kind of participatory real world gives rise to quantum states as rational states of belief?)} are strongly motivated by versions of this question. Arguing by analogy, we can characterize the situation in the structure terminology of Section~\ref{SecMath} as follows. Quantum mechanics, as it is used in actual scientific practice, corresponds to some structure $S$. Bohmians claim that the world \emph{actually} corresponds to a structure $S'$ that is more differentiated than $S$, carrying unverifiable\footnotemark\enspace answers to questions like ``where is the electron exactly, right now?''. Supporters of epistemic views in the sense above state as their goal to discover the correct differentiation $S'$.

The structure terminology suggests an obvious alternative: \emph{perhaps the question simply doesn't \textbf{have} an answer}. The urge to claim that it does, but that we are unable to find it, is arguably motivated by a metaphysics of ``things''. Similarly as our example of the natural numbers $\mathbb{N}$ in Section~\ref{SecMath}, such a view of the world depicts the universe as a collection of objects, or as a thing in itself, that ``sits there'' in an infinitely differentiated form. Similarly as a material body \emph{cannot not have a weight}, or a coin \emph{cannot not show heads or tails}, we tend to take it as an analytic truth that the world \emph{cannot not be in some configuration}. But if we see the world not as a thing, but as structure in some sense, then we may well accept the possibility that it corresponds more accurately to structure $S$ than to any of its differentiations:
\begin{shaded}
\textbf{Interpretation.} Modern physics has shown us that some apparent properties of the world are actually \emph{non-properties}: they correspond to questions that do not \emph{have} an answer.

While G\"odel's results are \emph{not} directly applicable to the physical world, they motivate a use of the \emph{structure terminology} to interpret this phenomenon by analogy. Structure manifests itself by, and weaves together, ``real patterns''~\cite{Dennett} (such as correlations in measured data). Structure can be more or less differentiated. \emph{Structural undifferentiation} means that there are questions that have no answers, or that there are less patterns than expected.
\end{shaded}
Apart from dissolving the question above, what else do we gain from a ``structural'' perspective? In some cases, the claim that a question doesn't have an answer can have surprising predictive power. As an example, consider \emph{device-independent quantum cryptography}~\cite{BHK}. Two agents (Alice and Bob) perform local measurements on entangled quantum states. They use the random, correlated outcomes to generate a secret key. Could there be an eavesdropper (Eve) that spies on their message? If Alice's and Bob's statistics violates a Bell inequality, then the answer must be ``no''. Namely, if Eve is constrained by locality, and the setup violates \emph{local realism}, then what remains is a phenomenal violation of realism: Eve cannot be correlated to any ``elements of reality'' in her past that would  correspond to that secret key. In other words: \emph{you cannot spy on something that doesn't exist}.
\addtocounter{footnote}{-1}
\footnotetext{{Note that this characterization does not apply to Valentini's version of de Broglie-Bohm theory~\cite{Valentini}, which includes the possibility to have \emph{quantum nonequilibrium systems} that make predictions which differ from standard quantum mechanics.}}

But the predictive power of this kind of reasoning seems to come at a price: aren't we giving up on \emph{realism} here?

\section{Ontic structural realism}
In the context of quantum physics, the word ``realism'' is ambiguous and overloaded. The violation of a Bell inequality implies a violation of \emph{local realism}, but these notions are defined in a very specific way. Interpretations of quantum theory that reject the violation of locality are often labelled as ``anti-realist''~\cite{Leifer}. But this includes interpretations of quantum mechanics that simply reject the mathematical framework on which the derivation is built in the first place (the ontological models framework~\cite{HarriganSpekkens}), even if they rest on a generally realist view of the physical world. A prominent example is given by QBism~\cite{Fuchs} that subscribes to a notion of ``participatory realism''.

In particular, to be a realist doesn't commit one to a metaphysics of ``things'' -- perhaps quite on the contrary. This is the main point in a book by Ladyman et al.\ with the title of the previous section: Every Thing Must Go~\cite{Ladyman}. The authors argue precisely for a form of realism relying on a metaphysics of structure, not things -- \emph{ontic structural realism}.

The goals of Ladyman et al.\ are quite different from those of this essay -- they are mostly motivated by the problems of standard scientific realism: \emph{``[...] the history of successful novel prediction science is the most compelling evidence for some form of realism, but [...] the history of ontological discontinuity across theory change makes standard scientific realism indefensible.''}

In particular, what is rejected is the `doctrine of containment': \emph{``On this doctrine, the world is a kind of container bearing objects that change location and properties over time. These objects cause things to happen by interacting directly with one another. [...] they themselves are containers in turn, and their properties and causal dispositions are to be explained by the properties and dispositions of the objects they contain (and which are often taken to comprise them entirely).''}

It is argued that this kind of view is in line with human intuition, but not so much with modern physics: \emph{``we should not interpret science [...] as metaphysically committed to the existence of self-subsistent individuals. [...] We will later say that what exists are (`real') `patterns'. [...] When we go on to deny that, strictly speaking, there are `things', we will mean to deny that in the material world as represented by the currently accepted scientific structures, individual objects have any distinctive status.''}

Such a version of realism is still able to account for the ``no-miracles argument''~\cite{Putnam}: that the best explanation of the success of science is that our best scientific theories are at least approximately true. In particular, it frees us from a problem famously described by Laudan~\cite{Laudan}: if we insist on understanding ``approximate truth'' as the property that the central terms of a scientific theory (such as Dalton's atoms or Bohr's electrons) refer to actual entities in the world, then we have to regard previous physical theories (and thus perhaps also contemporary ones) as utterly unsuccessful. On the other hand, if we base realism on a structural ontology of ``real patterns'', then this problem is dissolved, and a form of stability across theory change is established.

In summary: applying structural terminology to our understanding of the physical world is not in conflict with realism, but on the contrary implied by mature versions of it.

\section{Quantum-optimistic conclusions}
What can we conclude if we accept the structural view put forward in this essay? First, interpreting undecidability as undifferentiation of structure arguably renders ``anti-mechanist'' views as expressed for example by Driessen and Suarez~\cite{DriessenSuarez} implausible. \emph{Non-existing} answers can neither be found by machines, nor by humans -- nor by gods.

Regarding quantum theory, the structural perspective seems to bring us closer to views that are often unduly characterized as anti-realist: views in which quantum states are states of information about future measurement outcomes (or experiences), but not about some underlying reality~\cite{Leifer}. But we do not have to stop here. Seeing the world as consisting of real patterns interwoven by structure, not as a ``thing'' or a collection of things, opens up the possibility to reconcile these epistemic interpretations with others that regard the quantum state as ontic. Namely, if quantum states are expressions of knowledge (or belief, or chance), and if the quantum state ``is'' the world (or is in the world), then why not accept the conjunction of both views?
\begin{shaded}
\textbf{Hypothesis.} The quantum world \textbf{is} probabilistic structure. In other words, it is not a ``thing'' or a collection of things, but it is the multitude of statistical patterns and their structural relations that any observer encounters in their data.
\end{shaded}	
In~\cite{Law}, I have worked out a concrete version of this hypothesis in detail. It starts with the claim that the world \emph{is nothing but} the determination of the chances of what happens to any observer next, and derives our usual picture of a ``thing-like'' objective world from it. Regardless of this specific approach, the main message is one of \emph{optimism}: seeing unpredictability not as an expression of a fundamental epistemic restriction, but as structural undifferentiation admits new fruitful perspectives on the world, including ones that drop the false dichotomy of ``ontic'' and ``epistemic'' interpretations of the quantum state.

In this view, undecidability and unpredictability are not in themselves sufficient reasons to be pessimistic. But perhaps this is a dangerous perspective. As Steven Pinker~\cite{Pinker} points out,

\emph{Pessimism has been equated with moral seriousness.} Citing the popular naysayers, \emph{if you think knowledge can help solve problems, then you have a ``blind faith'' and a ``quasi-religious belief'' in the ``outmoded superstition'' and ``false promise'' of the ``myth'' of the ``onward march'' of ``inevitable progress''.}

In the light of G\"odel's theorems, the epitaph of David Hilbert's tombstone in G\"ottingen is sometimes regarded as a prototype of a false promise:

\textit{We must know. We will know.}

\noindent
Let me therefore conclude this essay with one further outmoded declaration of blind faith:\\

\textit{We can know what there is to know.}

\end{document}